\documentclass[aps,twocolumn,prd,showpacs,showkeys,preprintnumbers,superscriptaddress,nobibnotes,floatfix,longbibliography,nofootinbib]{revtex4-1}

\usepackage[T1]{fontenc} 
\usepackage[compat=1.1.0]{tikz-feynman}
\usepackage{pifont}
\usepackage[mathscr]{eucal}
\usepackage{multirow}
\usepackage{graphicx}
\usepackage{amsfonts}
\usepackage{amsmath}
\usepackage{amssymb}
\usepackage{epsfig}
\usepackage{slashed}
\usepackage{dcolumn}%
\usepackage{ulem}
\usepackage{color}

\usepackage[english]{babel}
\usepackage{microtype}
\usepackage{pifont}
\definecolor{myred}{RGB}{255, 0, 0}
\definecolor{myblue}{RGB}{0, 0, 255}
\definecolor{mygreen}{RGB}{0, 128, 0}

\usepackage[colorlinks,
linkcolor=blue,
citecolor=blue,
urlcolor=blue
]{hyperref}

\begin{document}
\title{The nano-hertz and milli-hertz stochastic gravitational waves in the minimal clockwork axion model}

\author{Xiangwei Yin}
\email{yinxiangwei@cqu.edu.cn}
\affiliation{Department of Physics and Chongqing Key Laboratory for Strongly Coupled Physics, Chongqing University, Chongqing 401331, P.R. China}

\author{Cheng-Wei Chiang}
\email{chengwei@phys.ntu.edu.tw}
\affiliation{Department of Physics and Center for Theoretical Physics, National Taiwan University, Taipei 10617, Taiwan}
\affiliation{Physics Division, National Center for Theoretical Sciences, Taipei 10617, Taiwan}

\author{Bo-Qiang Lu}
\email{bqlu@zjhu.edu.cn}
\affiliation{School of Science, Huzhou University, Huzhou, Zhejiang 313000, China}
\affiliation{Zhejiang Key Laboratory for Industrial Solid Waste Thermal Hydrolysis Technology and Intelligent Equipment}

\author{Tianjun Li}
\email{tli@itp.ac.cn}
\affiliation{CAS Key Laboratory of Theoretical Physics, Institute of Theoretical Physics,
Chinese Academy of Sciences, Beijing 100190, China}
\affiliation{School of Physical Sciences, University of Chinese Academy of Sciences,
No.~19A Yuquan Road, Beijing 100049, China}
\affiliation{School of Physics, Henan Normal University, Xinxiang 453007, P. R. China}

\begin{abstract}

The clockwork framework can realize TeV-scale $U(1)_{PQ}$ symmetry breaking while generating a large axion decay constant \(f_a\).
We propose a minimal clockwork axion model with three scalar fields, in which two domain walls (DWs) have non-zero tension. 
The DW associated with one of the fields is formed following the Peccei-Quinn (PQ) symmetry breaking and subsequently collapses due to the bias potential induced by the QCD instanton. The nano-hertz stochastic gravitational waves (GWs) generated from this DW annihilation can be probed by Pulsar Timing Arrays experiments.
In addition, the DW related to the other field is annihilated by a bias potential originating from higher-dimensional operators, producing a significant GW signal with a peak frequency around \(9.41\times10^{-5}\) Hz, which can be detected by the LISA, Taiji, and TianQin experiments.  Constraints on the model from SN1987, dark matter overproduction, Big Bang Nucleosynthesis, cosmic microwave background, and primordial black holes have been considered. The relic density of QCD axion dark matter can be explained through the misalignment mechanism. 

\end{abstract}
\maketitle

\newpage

\section{Introduction}

In quantum chromodynamics (QCD), the $U(1)$ problem~\cite{Weinberg:1975ui} emerges from the unobserved light pseudoscalar meson expected from the axial $U(1)$ symmetry in the Goldstone theorem, suggesting an anomaly in the conservation of the axial $U(1)$ current. 
The discovery of Yang-Mills instantons~\cite{Belavin:1975fg} and the non-trivial structure of the QCD vacuum~\cite{Callan:1976je,Jackiw:1976pf} have reshaped our understanding of QCD. These developments demonstrate that $U(1)_A$ does not constitute a true symmetry in QCD, thereby solving the $U(1)_A$ problem~\cite{tHooft:1976rip,tHooft:1976snw,tHooft:1986ooh}. Concurrently, these findings introduce the strong CP problem, as the non-trivial QCD vacuum implies the necessity of a CP-violating topological term, parametrized by the $\theta$ parameter. The neutron electric dipole moment (nEDM) emerges as one of the most sensitive probes of such CP violation. However, non-observation of CP violation in strong interactions deepens the mystery of the strong CP problem.
One of the most promising approaches to resolving the strong CP problem is to introduce a new global $U(1)_{PQ}$ symmetry, known as the Peccei-Quinn (PQ) symmetry~\cite{Peccei:1977hh,Peccei:1977ur}, which is first realized in a concrete model by Weinberg and Wilczek~\cite{Weinberg:1977ma,Wilczek:1977pj}.  As a result, a hypothetical pseudoscalar particle, the axion, would arise from the spontaneous breakdown of the $U(1)_{PQ}$ symmetry.
This particle offers a dynamic cancellation of the CP-violating term, effectively resolving the strong CP problem.

The traditional axion models require the axion decay constant to be in the range of $10^9 \mathrm{GeV} \lesssim f_a \lesssim 10^{12} \mathrm{GeV}$, which is of the same order as the scale of PQ symmetry breaking.  (See Ref.~\cite{DiLuzio:2020wdo} for a comprehensive review and also Ref.~\cite{Hook:2018dlk}, while the cosmological phenomenology can be found in Refs.~\cite{Marsh:2015xka,OHare:2024nmr}.)
It is an interesting question whether one can realize PQ symmetry breaking at the TeV scale while simultaneously generating a large axion decay constant $f_a$.  The clockwork axion model, first proposed in~\cite{Kaplan:2015fuy,Choi:2015fiu}, allows for a decoupling between the PQ symmetry breaking scale \(f_{\text{PQ}}\) and the axion decay constant $f_{a}$.
This alignment mechanism~\cite{Higaki:2016jjh,Higaki:2016yqk}, which requires the introduction of multiple axions, has its precursors described in the Kim-Nilles-Peloso (KNP) alignment~\cite{Kim:2004rp}, and was subsequently discussed in~\cite{Choi:2014rja}. 
For further phenomenological discussions, see Refs~\cite{Higaki:2015jag,Farina:2016tgd,Giudice:2016yja,Coy:2017yex,Long:2018nsl,Agrawal:2018mkd,Lu:2023mcz,Chiang:2020aui} and references therein.

The recent data released from NANOGrav ~\cite{NANOGrav:2023gor,NANOGrav:2023hvm}, along with EPTA~\cite{EPTA:2023fyk}, PPTA~\cite{Reardon:2023gzh}, and CPTA~\cite{Xu:2023wog}, present significant evidence for a Hellings-Downs pattern in angular correlations that is characteristic of gravitational waves (GWs). Notably, the NANOGrav 15-year (NG15) data provide the strongest constraints and the most compelling statistical support, reinforcing the existence of a nano-Hz stochastic gravitational wave background (GWB).

In the original clockwork axion model, the PQ symmetry-breaking scales corresponding to the scalar fields do not necessarily need to be identical. Motivated by this fact, we propose in this paper a minimal clockwork axion model consisting of three scalar fields, each with a distinct vacuum expectation value (VEV) \(f_i\) and PQ charge. As a result, the axion decay constant \(f_a\) receives enhancements from both the PQ charge and the VEV, bringing it within the traditional range.

We investigate the GWs signals generated by domain wall (DW) annihilation according to the model. 
The simulation results show that the scenario of DW annihilation induced by the QCD instanton effect requires at least three scalar fields~\cite{Higaki:2016jjh}, and therefore in this work we focus on this minimal field content. In the clockwork axion model, the DWs are formed by the additional pseudoscalar fields, referred to as gear fields $A_i$, with masses $m_{A_i}$ and tension $\sigma_i \simeq 8m_{A_i} f_{\mathrm{PQ}}^2$. For convenience, we denote the domain wall associated with each gear field $A_i$ as the $A_i$-wall.

In the canonical clockwork axion model, all scalar fields acquire the same VEV, and therefore the DWs associated with each gear field $A_i$ share the same annihilation mechanism, such as the instanton effect. In this work, however, since the VEVs exhibit a hierarchy, the annihilation mechanisms of the $A_i$-walls may not be identical. In the following, we assume that the $A_2$-wall annihilates via the instanton effect. This produces a GW signal with $\nu_{\text{peak}} \sim 10^{-7.5}~\mathrm{Hz}$ and $h^2\Omega_{\text{GW}}^{\text{peak}} \sim 10^{-6.4}$, providing a good fit to the recently announced NG15 data, with best-fit parameters $f_{2} = 10^{2.3}~\mathrm{TeV}$ and $\epsilon = 10^{-3}$. In contrast, the $A_1$-wall cannot be annihilated consistently by instanton effects. If one nevertheless assumes the instanton-driven annihilation, the calculated values $T_{\text{ann}} \simeq 0.078~\mathrm{MeV}$ and $T_{\text{dom}} \simeq 447.74~\mathrm{GeV}$ are inconsistent, as they imply that the $A_1$-wall would dominate the energy density of the Universe well before its annihilation. Therefore, the $A_1$-wall requires a different annihilation mechanism, such as a bias potential induced by higher-dimensional operators, which generates a second GW signal with $\nu_{\text{peak}}\simeq 9.41\times 10^{-5}~\mathrm{Hz}$ and $h^2\Omega_{\text{GW}}^{\text{peak}} \simeq 1.76\times 10^{-7}$. This predicted signal is detectable by the LISA~\cite{Caprini:2019egz}, Taiji~\cite{Hu:2017mde,Ruan:2018tsw}, and TianQin~\cite{TianQin:2015yph,TianQin:2020hid,Liang:2021bde}.
For completeness, in Appendix~\ref{AppendixA} we analyze the opposite case, namely that the $A_1$-wall annihilates via the instanton effect, while the $A_2$-wall requires an additional source of annihilation.
We have also considered the constraints on the model from SN1987, dark matter (DM) overproduction, Big Bang Nucleosynthesis (BBN), cosmic microwave background (CMB), and primordial black holes (PBH). The saturated relic density of QCD axion DM can be obtained via the misalignment mechanism.

This paper is organized as follows. In Sec.~\ref{Section2}, we introduce the minimal clockwork axion model. In Sec.~\ref{Section3}, we present the GW signals generated by DW annihilation and discuss the relevant experimental constraints. In Sec.~\ref{Section4}, we describe the fit to the NG15 data and provide further discussion on the DW annihilation scenario. We conclude in Sec.~\ref{Section5}.  Appendix~\ref{AppendixA} gives an alternative scenario to the one presented in the main text.

\section{The minimal clockwork axion model}\label{Section2}

The clockwork axion framework~\cite{Kaplan:2015fuy,Choi:2015fiu} introduces $N+1$ complex scalars, denoted by $\Phi_{i}(x)$ with $i=0,1,\dots, N$.  The associated renormalizable potential is given by
\begin{equation}
\resizebox{\linewidth}{!}{$
V(\Phi)=\sum_{i=0}^{N}\left(-m^2\left|\Phi_i\right|^2+\lambda\left|\Phi_i\right|^4\right)-\epsilon \sum_{i=0}^{N-1}\left(\Phi_i^{\dagger} \Phi_{i+1}^3+\text {H.c.}\right)
~,
$}
\label{Sec2_Eq_01}
\end{equation}
where the first term respects a $U(1)^{N+1}$ global symmetry, while the second term, involving the coupling $\epsilon$, explicitly breaks these $N+1$ global symmetries down to a single global $U(1)$ symmetry, which is identified as the PQ symmetry:
\begin{equation}
\mathrm{U}(1)_{PQ}: \Phi_i \rightarrow \exp \left[i q^{N-i} \theta\right] \Phi_i~,
\end{equation}
with $\theta \in[0,2 \pi)$ and $q\equiv3$.

The scalar field can be parametrized as 
\begin{equation}
    \Phi_i=\left(f_{i}+\rho_i\right) e^{i \pi_i / f_{i}} / \sqrt{2}~,
\end{equation}
where each $\rho_{i}$ and $\pi_i$ denote the radial and angular modes, respectively, and each field is equipped with a distinct VEV $f_{i}$.
In the absence of the $\epsilon$ term, the scalar field acquiring the above VEV would still break the $U(1)^{N+1}$ symmetry. However, in the presence of the $\epsilon$ term and after the spontaneous symmetry breaking, the particle spectrum contains $N$ massive particles and one massless particle. 
We are interested in the DWs as possible sources of GWs. The DW can survive until the QCD phase transition occurs when there are at least three fields~\cite{Higaki:2016jjh}. We therefore consider the scenario with this minimal field content.
The $U(1)_{PQ}$ charges and VEVs are given by
\begin{equation}
\begin{array}{rccc}
\hline & \Phi_0 & \Phi_1 & \Phi_2 \\
\hline U(1)_{PQ} & 3^2 & 3 & 1  \\
\hline \left\langle\Phi_i\right\rangle & f_0 & f_1 & f_2 \\
\hline
\end{array}
~,
\end{equation}
where the hierarchy in $f_{i}$ is $f_{0}=3^{5}f_{1}=3^{10}f_{2}$. The potential can now be written as
\begin{equation}
V(\pi)=-\frac{1}{2}\epsilon \sum_{i=0}^{1} f_{i}f_{i+1}^{3} \cos \left(3 \frac{\pi_{i+1}}{f_{i+1}}-\frac{\pi_i}{f_i}\right)~,
\end{equation}
with the corresponding mass matrix of the $\pi_i$ fields given by 
\begin{equation}
\left(\begin{array}{ccc}
\frac{f_1^3 \epsilon}{2 f_0} & -\frac{3 f_1^2 \epsilon}{2} & 0 \\
-\frac{3 f_1^2 \epsilon}{2} & \frac{9 f_0 f_1 \epsilon}{2}+\frac{f_2^3 \epsilon}{2 f_1} & -\frac{3 f_2^2 \epsilon}{2}  \\
 0 & -\frac{3 f_2^2 \epsilon}{2} & \frac{9 f_1 f_2 \epsilon}{2}
\end{array}\right)~.
\end{equation}
After diagonalizing the above mass matrix, we obtain two massive modes and one massless mode, corresponding to the QCD axion:
\begin{equation}
    m_a=0, m_{A_{1}}^{2}\simeq1093.5\epsilon f_{1}^{2}, m_{A_{2}}^{2}\simeq1093.5\epsilon f_{2}^{2}~.
\end{equation}
In this case, the tension of the DWs can be approximately expressed as $\sigma_{A_{i}} \simeq 8 m_{A_{i}} f_{i}^{2}$, where $i=0,1,2$ and $m_{A_{i}}$ is the mass of the $i$-th axion.

The key points are that the QCD axion $a$ mainly comes from $\Phi_{0}$ and that $f_{a}$ is enhanced by both PQ charge and VEV. 
Taking into account the fact that the QCD axion is expected to couple very weakly to gluons, we can couple $\Phi_{2}$ to a pair of vector-like quarks ($XQ$, $XQ^c$), analogous to the Kim-Shifman-Vainshtein-Zakharov (KSVZ)~\cite{Kim:1979if,Shifman:1979if} type axion model 
\begin{equation}
    \Delta \mathcal{L}=y \Phi_2 XQ^c XQ~,
\end{equation}
where the vector-like quarks  $XQ$ and $XQ^c$ have the Standard Model (SM) quantum numbers $(3,1,0)$ and $({\bar 3},1,0)$, respectively, as in the canonical KSVZ model. Its PQ charges are assigned as $\mathcal{X}_{{XQ}_L}=-\mathcal{X}_{{XQ}_R}=1/2$, which corresponds to the case with $N_{DW}=1$. After integrating out the heavy particles, the axion at low energies will couple to the topological term 
\begin{equation}
    \mathcal{L} \supset \frac{\alpha_s}{8 \pi} \frac{a}{f_a} G_{\mu \nu}^a \tilde{G}^{\mu \nu, a}~,
\end{equation}
with
\begin{equation}
    f_{a}= \sqrt{\sum_{0}^{2} q_{i}^{2} f_{i}^{2}} \simeq3^2 f_0=3^{7}f_{1}=3^{12}f_{2}~.
\end{equation}
In this study, we will consider the QCD axion as a dark matter candidate.

\section{Gravitational waves and experimental constraints }\label{Section3}

In this section, we will first discuss the general case of GW signals generated by DW annihilation and then apply it to the DWs in this minimal model.  We will also examine existing experimental constraints on the model.

\subsection{GWs from DW annihilation}

The DWs are associated with the spontaneous breaking of discrete symmetries.  In each causally disconnected Hubble patch, the field randomly settles into one of the degenerate minima. As the Universe evolves and multiple such patches re-enter the horizon after a few Hubble times, DWs form as field configurations that interpolate in space between neighboring vacua.
The evolution of the energy density of DWs follows the so-called scaling solution~\cite{Press:1989yh,Garagounis:2002kt,Oliveira:2004he,Avelino:2005kn,Leite:2011sc,Leite:2012vn,Martins:2016ois,Hindmarsh:1996xv,Hindmarsh:2002bq}, and the energy density scales as
\begin{equation}
\rho_{\text{w}}(t)=\mathcal{A} \frac{\sigma}{t}~,
\label{rho_wall}
\end{equation}
where $\mathcal{A}\simeq 0.8 \pm 0.1$~\cite{Hiramatsu:2012sc}, $\sigma$ is the surface tension, and $t$ is the cosmic time.
For DWs, the tension force $p_{T}$ is equivalent to $\rho_{\text{w}}$ and the volume pressure force on the wall is estimated as $p_V \sim V_{\text{bias}}$. The QCD instanton effects cause a bias in the potential $V_{\text{bias}}\sim \Lambda_{\text{QCD}}^{4}$, which leads to the annihilation of the DWs when $p_{V}\sim p_{T}$.
Therefore, the annihilation time can be estimates as~\cite{Saikawa:2017hiv}~\footnote{see~\cite{Babichev:2025stm} for more discussion on $t_{\text{ann}}$}
\begin{equation}
\begin{aligned}
t_{\text {ann }} 
& =
C_{\text {ann }} \frac{\mathcal{A} \sigma}{V_{\text {bias }}}
\\
& =
6.58 \times 10^{-4}~\mathrm{s} ~\mathrm{C}_{\text {ann }} \mathcal{A}\left(\frac{\sigma}{\mathrm{TeV}^3}\right)\left(\frac{V_{\text {bias }}}{\mathrm{MeV}^4}\right)^{-1}
~,
\end{aligned}
\label{tann_00}
\end{equation}
where $C_{\text{ann}}$ is an $\mathcal{O}(1)$ coefficient. If the annihilation occurs during the radiation dominated era, Eq.~(\ref{tann_00}) can be rewritten as
\begin{equation}
\begin{aligned}
T_{\text {ann }} 
\simeq & 
1.35 \times 10^{-2} \mathrm{GeV} \epsilon^{-1 / 4}\left(\frac{g_*\left(T_{\text {ann }}\right)}{10}\right)^{-1 / 4}
\\
& \times\left(\frac{f}{100~ \mathrm{TeV}}\right)^{-3 / 2}\left(\frac{\Lambda_{\mathrm{QCD}}}{100~ \mathrm{MeV}}\right)^2
~,
\end{aligned}
\label{Tann_01}
\end{equation}
where we take $C_{\text{ann}}\simeq 3$, $\mathcal{A}\simeq 0.8$, $t=\frac{1}{2H}$, $H^{2}=\frac{8\pi G}{3}\rho_{r}$, $\rho_{r}=\frac{\pi^{2}}{30}g_{*}(T_{\text{ann}})T_{\text{ann}}^{4}$, and $\sigma\simeq 8 m f^2$.

According to Ref.~\cite{Saikawa:2017hiv}, GWs are produced during the annihilation of DWs, with the peak amplitude occurring at the moment of their annihilation:
\begin{equation}
\Omega_{\mathrm{GW}}^{\text {peak }}\left(t_{\mathrm{ann}}\right)=\frac{8 \pi \tilde{\epsilon}_{\mathrm{GW}} G^2 \mathcal{A}^2 \sigma^2}{3 H^2\left(t_{\mathrm{ann}}\right)}~,
\end{equation}
where $\tilde{\epsilon}_{\text{GW}}$ is estimated as $\tilde{\epsilon}_{\mathrm{GW}} \simeq 0.7 \pm 0.4$~\cite{Hiramatsu:2013qaa} and $\mathcal{A}\simeq N$~\cite{Higaki:2016jjh} from two-dimensional lattice simulations.
The peak amplitude of GWs at the present time is then given by~\cite{Saikawa:2017hiv}
\begin{equation}
\begin{split}
h^2\Omega_{\mathrm{GW}}^{\text {peak }}\left(t_0\right)
=&
7.2 \times 10^{-18} \tilde{\epsilon}_{\mathrm{GW}} \mathcal{A}^2\left(\frac{g_{* s}\left(T_{\mathrm{ann}}\right)}{10}\right)^{-4 / 3}
\\
&
\times \left(\frac{\sigma}{1~ \mathrm{TeV}^3}\right)^2\left(\frac{T_{\mathrm{ann}}}{10^{-2} ~\mathrm{GeV}}\right)^{-4} 
~,
\end{split}
\label{omega_peak}
\end{equation}
where $h$ is the reduced Hubble parameter, and the peak frequency is
\begin{equation}
\begin{aligned}
\nu_{\text {peak }}\left(t_0\right) 
\simeq & 
1.1 \times 10^{-9} \mathrm{~Hz}\left(\frac{g_*\left(T_{\text {ann }}\right)}{10}\right)^{1 / 2}
\\
&
\times \left(\frac{g_{* s}\left(T_{\text {ann }}\right)}{10}\right)^{-1 / 3}\left(\frac{T_{\text {ann }}}{10^{-2}~ \mathrm{GeV}}\right)
~.
\end{aligned}
\label{nu_peak}
\end{equation}

As alluded to before, the DW tension is given by $\sigma_{A_{i}} \approx 8 m_{A_{i}} f_{i}^{2}$. We refer to the DW associated with $A_i$ as the $A_i$-wall. In the following, we focus on the properties of the $A_2$-wall, while the phenomenology related to the $A_1$-wall will be discussed in the next section.
In the original clockwork axion model, all PQ symmetry-breaking scales are identical, leading to massive axions with nearly degenerate masses and DWs with approximately equal tensions. In contrast, the DWs in the minimal clockwork axion model proposed in this work exhibit significantly different tensions. 
The \(A_2\)-wall has a tension of $\sigma_{A_{2}} = 8m_{A_{2}}f_{2}^{2}$.  According to Eq.~(\ref{omega_peak}), the corresponding peak amplitude of GWs in this case is given by
\begin{equation}
\begin{aligned}
h^2 \Omega_{\mathrm{GW}}^{\text {peak}}\left(t_0\right) 
=&
4.53\times 10^{-4} \epsilon \tilde{\epsilon}_{\mathrm{GW}} \left(\frac{N}{3}\right)^2\left(\frac{g_{* s}\left(T_{\mathrm{ann}}\right)}{10}\right)^{-4 / 3}
\\
&
 \times\left(\frac{f_{2}}{100~ \mathrm{TeV}}\right)^6\left(\frac{T_{\mathrm{ann}}}{0.1~ \mathrm{GeV}}\right)^{-4}
 ~,
\end{aligned}
\label{omega_peak_min}
\end{equation}
and the GW spectrum can be scaled as
\begin{equation}
h^2 \Omega_{\mathrm{GW}}
= 
\begin{cases}
\displaystyle
h^2 \Omega_{\mathrm{GW}}^{\text {peak }}\left(\frac{\nu}{\nu_{\text {peak }}}\right)^3 , & \text { for } \nu<\nu_{\text {peak }} ,
\\ 
\displaystyle
h^2 \Omega_{\mathrm{GW}}^{\text {peak }}\left(\frac{\nu_{\text {peak }}}{\nu}\right) , & \text { for } \nu>\nu_{\text {peak }} .
\end{cases}
\label{shape_function}
\end{equation}

\subsection{Experimental constraints}

In this subsection, we discuss constraints on the model from various existing observations.

\textbf{Constraint from avoiding DW domination:}
In the scaling regime, the energy density of DWs decays as $\rho_{\text{w}} \propto t^{-1}$, which is slower than the decays of matter $\rho_{\text{m}} \propto t^{-3/2}$ and radiation $\rho_{\text{r}} \propto t^{-2}$. Therefore, the energy density of the Universe will be dominated by the DWs after $\rho_{\text{w}}(t_{\text{dom}})=\rho_{\text{c}}(t_{\text{dom}})$ and the corresponding temperature is
\begin{equation}
T_{\mathrm{dom}}
=
0.235 \mathrm{GeV} \epsilon^{1 / 4} \left(\frac{g_*\left(T_{\mathrm{dom}}\right)}{10}\right)^{-1 / 4} \left(\frac{f_2}{100~ \mathrm{TeV}}\right)^{3 / 2}~.
\label{Tdom}
\end{equation}
To ensure that the DWs annihilate before they dominate the energy density of the Universe $t_{\text{ann}} \lesssim t_{\text{dom}}$, we have the following constraint
\begin{equation}
f_2 
\lesssim 
38.6 ~ \mathrm{TeV} \epsilon^{-1 / 6}\left(\frac{g_*\left(T_{\mathrm{ann}}\right)}{g_*\left(T_{\mathrm{dom}}\right)}\right)^{-1/12} \left(\frac{\Lambda_{\mathrm{QCD}}}{100 ~\mathrm{MeV}}\right)^{2 / 3}
~.
\label{TannTdom}
\end{equation}

\textbf{Constraints from SN 1987A and oversaturated DM relic density:}
The axion decay constant $10^{9}~\text{GeV}\lesssim f_{a}\lesssim 10^{12}~\text{GeV}$ should be fulfilled.
The lower bound is derived from the SN 1987A neutrino burst observations~\cite{Mayle:1987as,Raffelt:1987yt,Turner:1987by}, while the upper bound is imposed to prevent the axion DM from overclosing the Universe, avoiding the need for fine-tuning the initial misalignment angle $\theta_i$~\cite{Preskill:1982cy,Abbott:1982af,Dine:1982ah}, which would otherwise invoke the anthropic axion.

\textbf{Constraint from BBN:}
Although we have no concrete knowledge of events prior to BBN, the thermal history after BBN is strictly constrained. We impose the conservative condition $T_{\mathrm{ann}}>T_{\mathrm{BBN}} \simeq 5 \mathrm{MeV}$, which corresponds to
\begin{equation}
    f_{2}
    \lesssim 
    193 ~\mathrm{TeV}\epsilon^{-1/6} \left(\frac{g_*\left(T_{\text {ann }}\right)}{10}\right)^{-1 / 6} \left(\frac{\Lambda_{\mathrm{QCD}}}{100~ \mathrm{MeV}}\right)^{4/3}
    ~.
\end{equation}

\textbf{Constraint from CMB:}
The relic density of relativistic axions is characterized by the effective number of relativistic species, $\Delta N_{\mathrm{eff}}$, which encodes the influence of dark radiation. Under the assumption of adiabatic expansion from the QCD epoch to recombination, we have~\cite{Long:2018nsl}
\begin{equation}
    \Delta N_{\text {eff }} 
    \simeq 
    0.298 v_w^2 \epsilon^{1 / 2}(\frac{f_{2}}{ 100~\mathrm{TeV}})^3
    ~,
\end{equation}
where $v_{w}\simeq 1$ is effective oscillation velocity of the wall. The current Planck (TT, TE, EE+lowE+lensing+BAO) constraints place an upper bound of $\Delta N_{\mathrm{eff}}<0.30$~\cite{Planck:2018vyg} at the $95\%$ confidence level (CL). The CMB-S4 will constrain $\Delta N_{\mathrm{eff}} \leq 0.06$~\cite{Abazajian:2019eic} at the $95\%$ CL.
The present spectrum, expressed in terms of the effective number of relativistic species, satisfies \( h^2 \Omega_{\mathrm{GW}}^{\mathrm{peak}}(t_0) \lesssim 5.6 \times 10^{-6} \Delta N_{\mathrm{eff}} \)~\cite{Caprini:2018mtu}.

\textbf{Constraint from PBHs:}
Primordial black holes form when closed DWs collapse after shrinking below their Schwarzschild radius~\cite{Ferrer:2018uiu,Gouttenoire:2023ftk,Gouttenoire:2023gbn}
\begin{equation}
    R\left(t_{\mathrm{PBH}}\right)
    =
    2 G M\left(t_{\mathrm{PBH}}\right)
    ~.
\end{equation}
Numerical results indicate that when accounting for $M\left(t_{\mathrm{PBH}}\right)$ contributions from the so-called bulk, boundary, and binding terms and under the condition that the Schwarzschild radius matches the Hubble horizon, the approximate relation $t_{\mathrm{PBH}}\simeq t_{\mathrm{dom}}$ can be obtained. In order to prevent PBHs from forming before the DW annihilation, the condition $t_{\mathrm{PBH}} \gtrsim t_{\mathrm{ann}}$, which coincides with the constraint previously obtained in Eq.~(\ref{TannTdom}), is naturally fulfilled.

\section{Data analysis and results}\label{Section4}

In this section, we examine the GWs produced by the $A_2$-wall annihilation driven by the QCD instanton effect, and investigate how they can fit the NG15 data and yield the corresponding best-fit parameters.  We then discuss the annihilation of the $A_1$-wall.  Since $\epsilon$ is universal and fixed by the $A_2$-wall, we will show that the $A_1$-wall requires another annihilation mechanism.

\subsection{Explanation of NG15 data by \(A_2\)-wall annihilation} 

The GWs produced by the \(A_2\)-wall annihilation in the minimal clockwork axion model can be expressed as a function of $f_2$ and $\epsilon$, as described by Eq.~(\ref{omega_peak_min}) and Eq.~(\ref{shape_function}). We adopt this signal as a template to fit the NG15 dataset~\cite{NANOGrav:2023gor}. The analysis is performed using {\tt PTArcade}~\cite{Mitridate:2023oar} code, a wrapper of {\tt ceffyl}~\cite{Lamb:2023jls} and {\tt ENTERPRISE}~\cite{Ellis:2020} to implement Bayesian inference based on pulsar timing data. {\tt PTArcade} samples the posterior distributions of the model parameters by employing the Markov Chain Monte Carlo (MCMC) techniques, allowing us to constrain $\epsilon$ and $f_2$ through a fit to the observed data.
By fixing $N=3$, we scan over $\epsilon$ and $f_2$ to obtain their posterior distributions.

\begin{figure}[htbp]
    \centering
    \includegraphics[width=\linewidth]{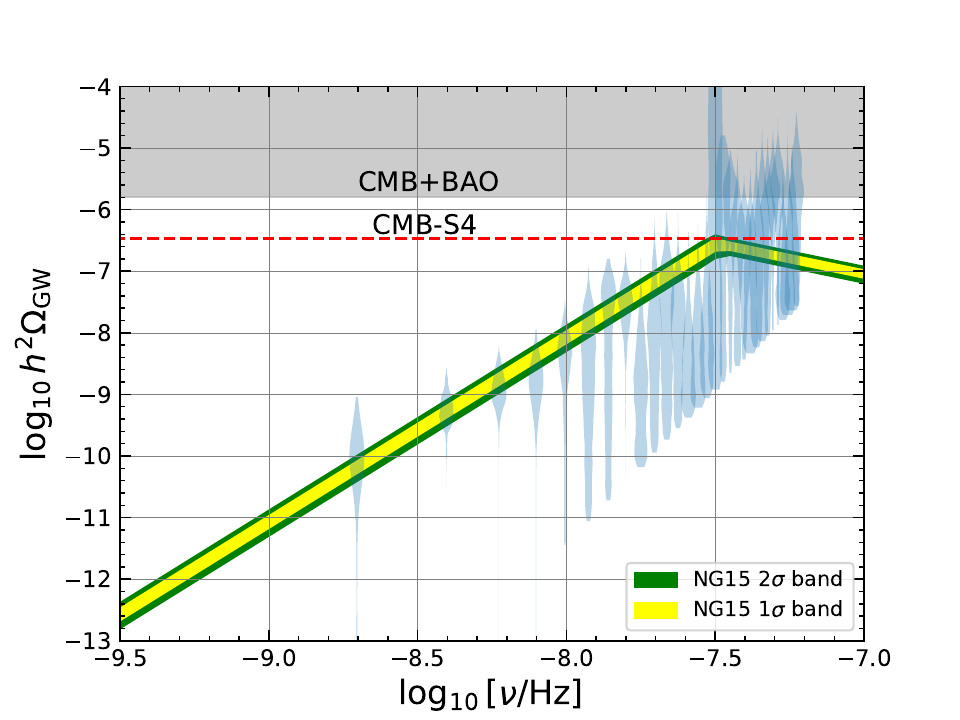}
    \caption{The GW signal spectrum versus frequency in the minimal clockwork axion model, originating from the $A_2$-wall annihilation. The yellow and green regions correspond to the $1 \sigma$ and $2 \sigma$ bounds derived from fitting the NG15 data, respectively. The blue violins give the posterior distributions of the free GW spectrum obtained from the NG15 data.}
    \label{IV_fig_01}
\end{figure}

In Fig.~\ref{IV_fig_01}, the yellow and green regions correspond to the $1 \sigma$ and $2 \sigma$ bounds obtained from fitting the NG15 data, respectively. The blue violins represent the posterior distributions of the free GW spectrum.
Following the NG15 analysis, we restrict our fit to the first 14 frequency bins (approximately 2–28~nHz) to avoid possible contamination from excess white noise at higher frequencies, ensuring consistency with the frequency range where the pulsar timing data provide the most reliable signal-to-noise ratio. Notably, our model predicts that the $A_2$-wall-induced GW spectrum continues to grow slightly beyond this range, reaching a peak frequency around $10^{-7.5}$~Hz with a corresponding amplitude of $10^{-6.4}$.
The peak of this signal lies beyond the sensitivity of the current Planck experiment~\cite{Planck:2018vyg}, as indicated by the gray region, but is within the detection capabilities of future CMB-S4~\cite{Abazajian:2019eic}, as shown with the red dashed line.

\begin{figure}[htbp]
    \centering
    \includegraphics[width=\linewidth]{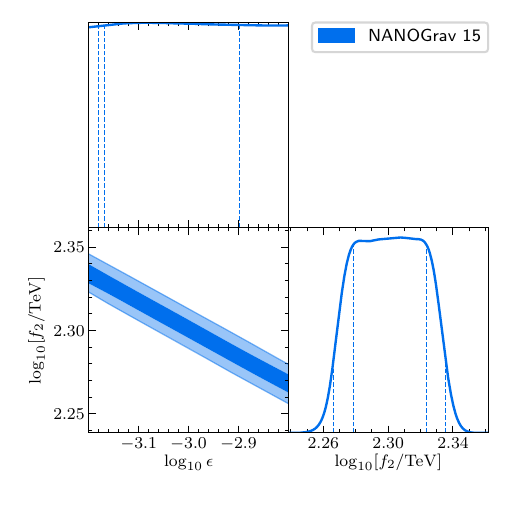}
    \caption{Constraints from NG15~\cite{NANOGrav:2023gor} on the parameters of GWs generated by $A_{2}$-wall annihilation in the minimal clockwork axion model. The best-fit values are \(f_{2} = 10^{2.3} \, \mathrm{TeV}\) and \(\epsilon = 10^{-3}\).}
    \label{IV_fig_02}
\end{figure}

Figure.~\ref{IV_fig_02}, generated using {\tt GetDist}~\cite{Lewis:2019xzd}, presents the fitting results for the model parameters in the minimal clockwork axion model, with best-fit values of \(f_{2} = 10^{2.3} \, \mathrm{TeV}\) and \(\epsilon = 10^{-3}\).
This set of best-fit parameters satisfies the previously discussed constraints from avoiding DW domination, SN 1987A, oversaturated DM relic density, BBN, CMB, and PBHs.
In this case, we obtain an axion decay constant \(f_{a} \simeq 1.1 \times 10^{11} \, \mathrm{GeV}\), which is enhanced by both the PQ charge and the VEV of the scalar fields to fall within the conventional range of the axion decay constant.
Consequently, the correct relic density can be obtained via the misalignment mechanism, with $\Omega_a h^2 = 0.01\, \theta_0^2\, \left(\frac{f_a}{10^{11} \, \mathrm{GeV}}\right)^{1.19}$ with $\theta_0\sim \mathcal{O}(1)$~\cite{Hook:2018dlk}.
This minimal framework stands in contrast to prior studies~\cite{Lu:2023mcz}, notably allowing a viable parameter space for $N=3$, which is typically ruled out in conventional clockwork axion constructions.

\subsection{Consequences of \(A_1\)-wall}

As mentioned previously, the DWs should annihilate before they dominate the energy density of the Universe. Therefore, we need to examine whether the \(A_1\)-wall would lead to the overclosure of the Universe.
One may be tempted to use Eq.~(\ref{Tann_01}) and Eq.~(\ref{Tdom}) to calculate the annihilation and domination temperature of the \(A_1\)-wall with the best-fit \(\epsilon=10^{-3}\), thereby obtaining \(T_{\text{ann}} \simeq 0.078 \, \text{MeV}\) and \(T_{\text{dom}} \simeq 447.74 \, \text{GeV}\). 
Clearly, this naive estimate is incorrect,  
since the $A_1$-wall would dominate the energy density of the Universe well before it annihilates, which is inconsistent and implies that the $A_1$-wall cannot annihilate through instanton effects.  Therefore, a different annihilation mechanism is needed.

\begin{figure}[htbp]
    \centering
    \includegraphics[width=\linewidth]{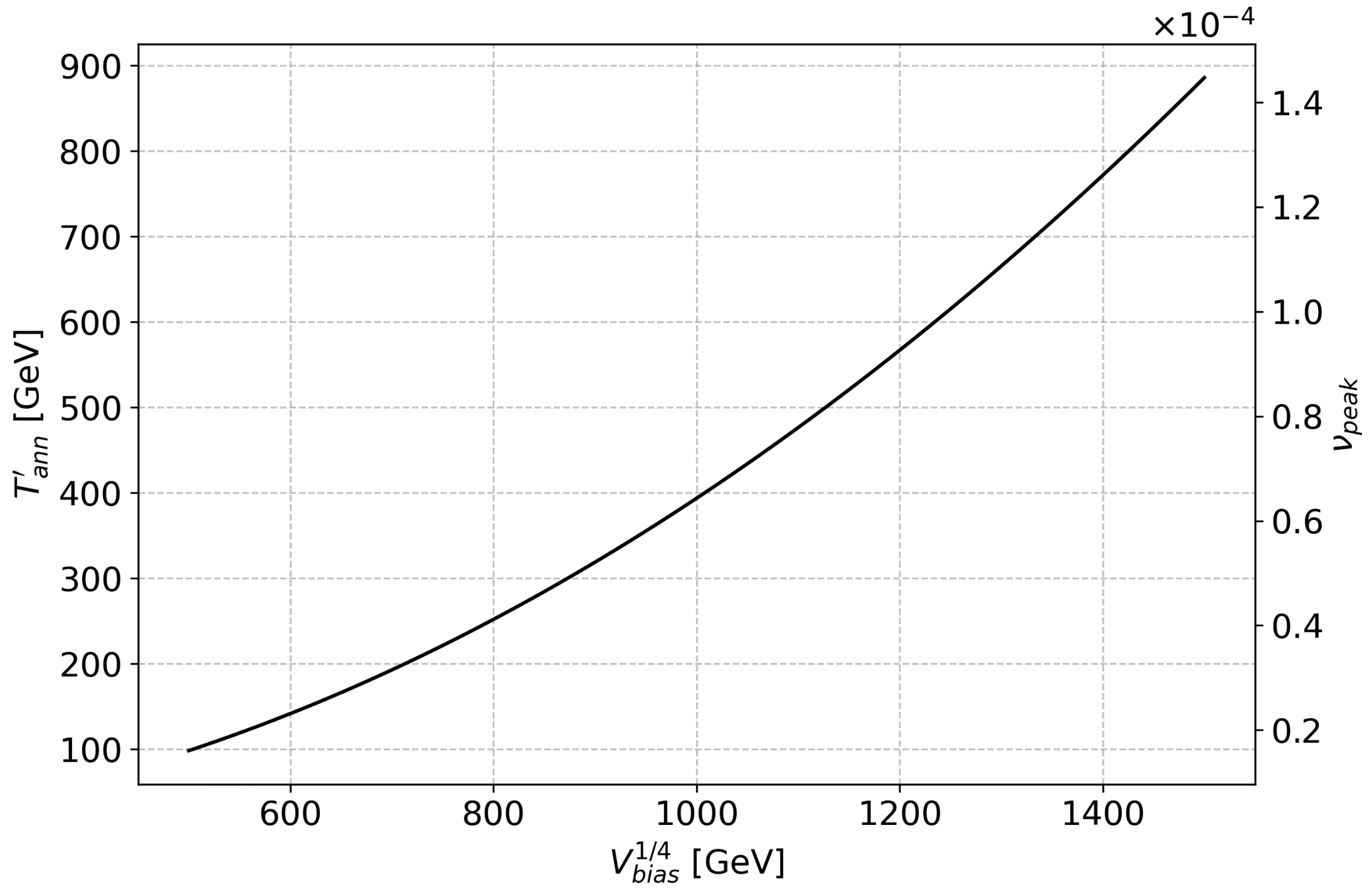}
    \sloppy
    \caption{Annihilation temperature $T_{\mathrm{ann}}^{\prime}$ and peak frequency $\nu_{\mathrm{peak}}$ as a function of bias potential. We use $\sigma_{w}\simeq8m_{A_1}f_1^2$, with $\epsilon=10^{-3}$, $f_1=3^{5}f_{2}$, and $f_2=10^{2.3} \text{TeV}$.  }
    \label{IV_fig_Tann_nu_Vbias}
\end{figure}

This problem can be trivially circumvented by introducing a field-dependent $\epsilon_{i}$ in Eq.~(\ref{Sec2_Eq_01}). Alternatively, we can consider a new bias term arising from higher-dimensional operators to annihilate the \(A_{1}\)-wall.
The annihilation occurs when the bias energy is comparable to the surface energy, and the general annihilation temperature derived from Eq.~(\ref{tann_00}) is
\begin{equation}
\begin{aligned}
    T_{\mathrm{ann}}^{\prime} 
    \simeq &
    110.5~\mathrm{MeV}\left(\frac{1}{\mathcal{A}}\right)^{1 / 2}\left(\frac{100}{g_*\left(T_{\mathrm{ann}}\right)}\right)^{1 / 4}\left(\frac{V_{\text {bias }}^{1 / 4}}{0.1 \mathrm{GeV}}\right)^2 
    \\
    &
    \times \left(\frac{10^5 \mathrm{GeV}}{\sigma_w^{1 / 3}}\right)^{3 / 2}
    \label{Gen_Tann}
\end{aligned}
~,
\end{equation}
where we take $C_{\mathrm{ann}}\simeq 3$. Figure~\ref{IV_fig_Tann_nu_Vbias} illustrates the annihilation temperature \(T_{\text{ann}}^{\prime}\) and peak frequency \(\nu_{\text{peak}}\) as a function of the bias potential.  Using the parameter $\epsilon=10^{-3}$ fixed by the \(A_2\)-wall will lead to inconsistencies. 
To address this issue, a bias potential can be introduced. For instance, by setting $V_{\text {bias }}^{1 / 4}=1200$~GeV, we obtain \( T_{\text{ann}}^{\prime} \simeq 576.5\ \mathrm{GeV} \) and $\nu_{\text{peak}}\simeq 9.41\times 10^{-5}$~Hz from Eq.~(\ref{nu_peak}).

\begin{figure}[htbp]
    \centering
    \includegraphics[width=\linewidth]{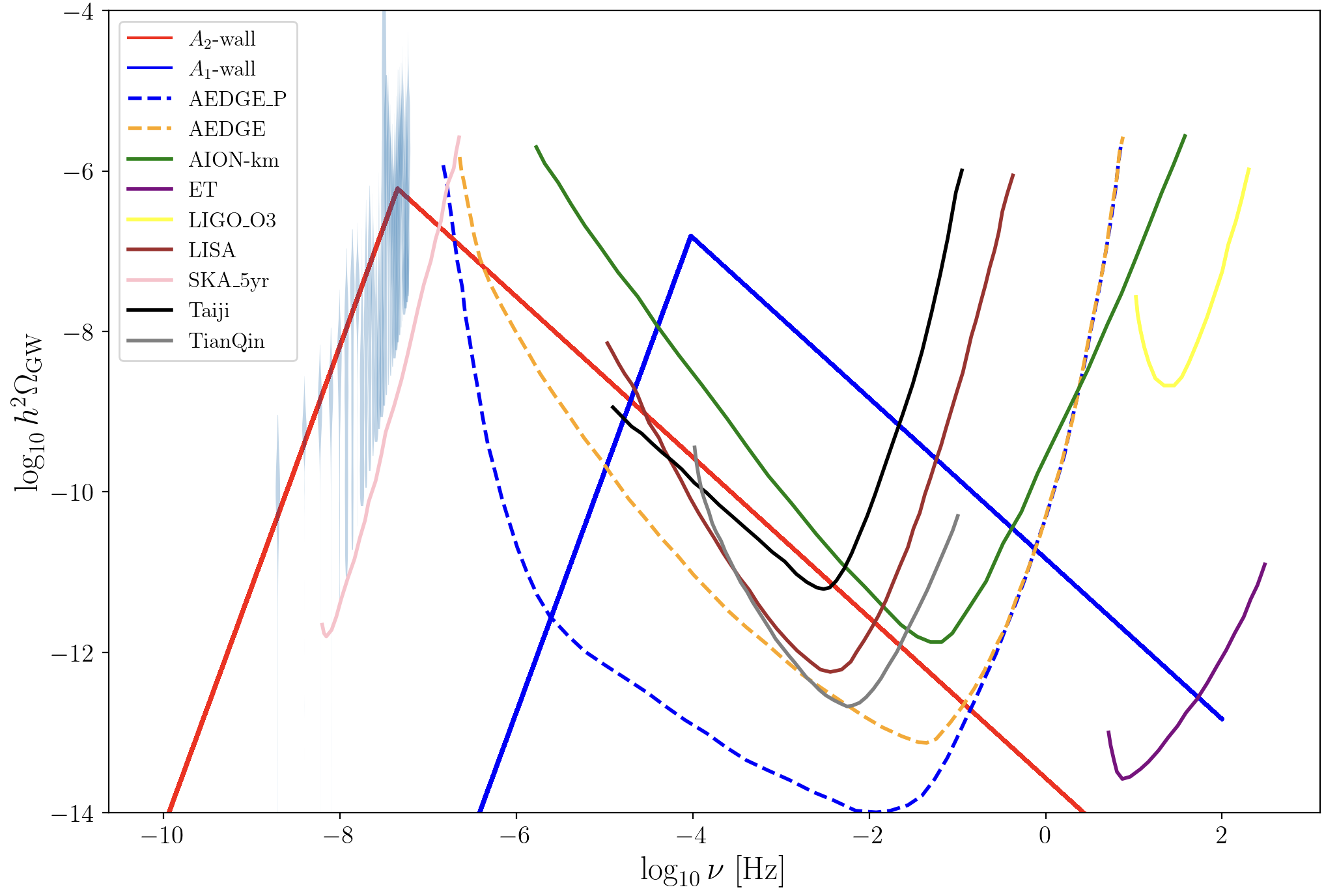}
    \caption{Two GW signals in the minimal clockwork axion model. The red ($\epsilon = 10^{-3}$ and $f_2 = 10^{2.3}\,\text{TeV}$) and blue ($\epsilon = 10^{-3}$, $f_1 = 3^5 f_2$, and $V_{\text{bias}}^{1/4} = 1200\,\text{GeV}$) lines represent the GW signals induced by the annihilation of the \(A_2\)-wall and the \(A_1\)-wall, respectively. In the plot, we also superimpose the sensitivity curves of SKA 5yr~\cite{Janssen:2014dka}, AION-km~\cite{Badurina:2019hst}, AEDGE~\cite{AEDGE:2019nxb}, AEDGE+, LISA~\cite{Caprini:2019egz}, LIGO O3~\cite{Shoemaker:2019bqt}, Einstein telescope (ET)~\cite{Maggiore:2019uih}, Taiji~\cite{Hu:2017mde,Ruan:2018tsw}, and TianQin~\cite{TianQin:2015yph,TianQin:2020hid,Liang:2021bde} experiments. }
    \label{IV_fig_03}
\end{figure}

The value $\Omega_{\text{GW}}^{\text{peak}}\simeq 1.76\times 10^{-7}$ can be obtained from Eq.~(\ref{omega_peak}), under the assumption that $g_{*}(T_\mathrm{ann})=g_{*s}(T_\mathrm{ann})=106.75$.
Eventually, two GW signals are identified, as illustrated in Fig.~\ref{IV_fig_03}. The GWs associated with the \(A_2\)-wall (red) has a peak frequency of \(\nu_{\text{peak}} \simeq 10^{-7.5}\)~Hz, while the signal originating from the \(A_1\)-wall (blue) exhibits a peak frequency of \(\nu_{\text{peak}} \simeq 9.41 \times 10^{-5}\)~Hz, making it detectable by the LISA~\cite{Caprini:2019egz}, Taiji~\cite{Hu:2017mde,Ruan:2018tsw}, and TianQin~\cite{TianQin:2015yph,TianQin:2020hid,Liang:2021bde} experiments.

We note in passing that even with the introduction of higher-dimensional operators, the strong CP phase can still remain sufficiently small.
For example, one may introduce the operator $V_{\text{PQ-break}}=\lambda_i e^{-i\delta_i}\Phi_i^5 / m_{\text{Pl}}$~\cite{DiLuzio:2020wdo}, which explicitly breaks the PQ symmetry. The bias potential $V_{\text{bias}}$ is controlled by the parameter $\lambda_1$, as $V_{\text{bias}}^{1/4} \propto V_{\text{PQ-break}}^{1/4}$. Meanwhile, the effective CP-violating phase $\bar{\theta}$ can be adjusted via the phase $\delta_1$, following $\langle\bar{\theta}\rangle\propto \frac{m_*^2 \tan \delta_1}{m_a^2+m_*^2}$, where $m_*^2=\frac{\lambda f_{1}^{2}}{2}(f_1/(\sqrt{2}m_{Pl}))\cos\delta_1$. Although some degree of fine-tuning is required for both $\lambda_1$ and $\delta_1$, this provides a viable mechanism for addressing the issue. Meanwhile, since $\lambda_2$ and $\delta_2$ can be independently controlled, the phenomenology associated with the $A_2$-wall, specifically the pattern of the potential bias induced by instanton effects, remains unaffected.
An alternative approach to circumvent the $A_1$-wall annihilation problem is to introduce the field-dependent $\epsilon_{i}$ in Eq.~(\ref{Sec2_Eq_01}).

\section{conclusion}\label{Section5}

The clockwork mechanism achieves a separation of the \(U(1)_{PQ}\) symmetry breaking scale \(f_{PQ}\) and the axion decay constant \(f_{a}\).
In this work, we propose a minimal clockwork axion model consisting of three scalar fields, each with a distinct vacuum expectation value and PQ charge. Consequently, the axion decay constant \(f_a\) is enhanced by both the PQ charges and the vacuum expectation values \(f_{i}\) of the scalar fields.
In this minimal model, there are three fields corresponding to the QCD axion and two massive axion fields, also referred to as the gear fields. 
We have investigated the GW signals produced by DW annihilation and identified two distinct GW signals.
The \(A_2\)-wall, annihilated by a potential bias from QCD instanton effects, generates nano-hertz GW signal that can explain the NG15 data, with the best-fit parameters \(f_2 = 10^{2.3} \, \mathrm{TeV}\) and \(\epsilon = 10^{-3}\).
However, the \(A_1\)-wall cannot annihilate via the same QCD instanton effects. To address this issue, we consider a bias potential arising from higher-dimensional operators, which enables DW annihilation and generates a prominent milli-hertz GW signal that can be probed by the LISA, Taiji, and TianQin experiments.
Additionally, the minimal model successfully satisfies constraints from SN1987, DM overproduction, BBN, CMB, and PBHs, while accounting for the saturated relic density of QCD axion DM via the misalignment mechanism. 
Our findings highlight the potential of the clockwork axion model in explaining axion physics and predicting detectable GW signals across multiple frequency ranges.

\acknowledgments

This research is supported in part by the National
Key Research and Development Program of China Grant No. 2020YFC2201504, by the
Projects No. 11875062, No. 11947302, No. 12047503, No. 12275333, and No. 12347101 supported by
the National Natural Science Foundation of China, by the Key Research Program of the
Chinese Academy of Sciences, Grant No. XDPB15, by the Scientific Instrument Developing
Project of the Chinese Academy of Sciences, Grant No. YJKYYQ20190049, and by the
International Partnership Program of Chinese Academy of Sciences for Grand Challenges,
Grant No. 112311KYSB20210012, and in part by the National Science and Technology Council under Grant Nos. NSTC-111-2112-M-002-018-MY3 and 114-2112-M-002-020-MY3, by the National Natural Science Foundation of China under Grant No.~12405058, and by the Zhejiang Provincial Natural Science Foundation of China under Grant No.~LQ23A050002.  This work was also performed in part at the Aspen Center for Physics, which is supported by a grant from the Simons Foundation (1161654, Troyer).

\appendix
\section{An alternative scenario}\label{AppendixA}

\subsection{Explanation of NG15 data by \(A_1\)-wall annihilation}

In the main text, we identified an inconsistency that the \(A_2\)-wall annihilates through QCD instanton-induced bias term while the \(A_1\)-wall requires a distinct annihilation mechanism. The key point is that \(\epsilon\) and \(f\) have a one-to-one correspondence, meaning that the NG15 data tied to \(f_2\) fixes the value \(\epsilon=10^{-3}\), preventing the use of identical parameters to describe the phenomenology related to \(A_1\)-wall.

\begin{figure}[htbp]
    \sloppy
    \centering
    \includegraphics[width=\linewidth]{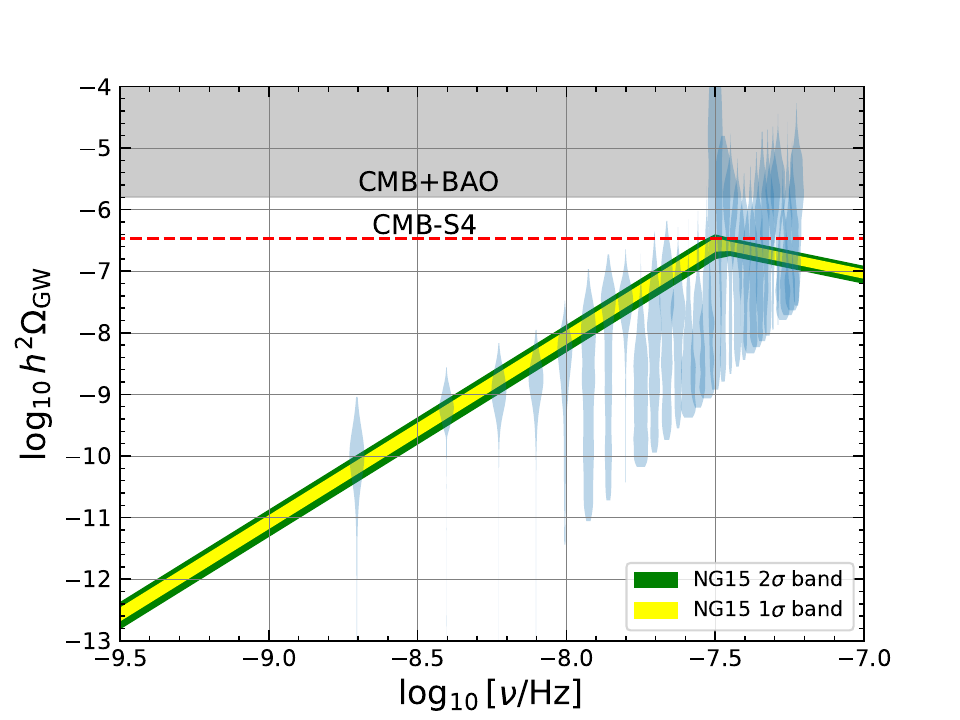}
    \caption{The GW signal versus frequency in the minimal clockwork model, originating from the $A_1$-wall annihilation. The yellow and green regions correspond to the $1 \sigma$ and $2 \sigma$ bounds derived from fitting the NG15 data, respectively. The blue violins represent the posterior distributions of the free GW spectrum obtained from the NG15 data.}
    \label{A_fig_01}
\end{figure}

\begin{figure}[htbp]
    \centering
    \includegraphics[width=0.9\linewidth]{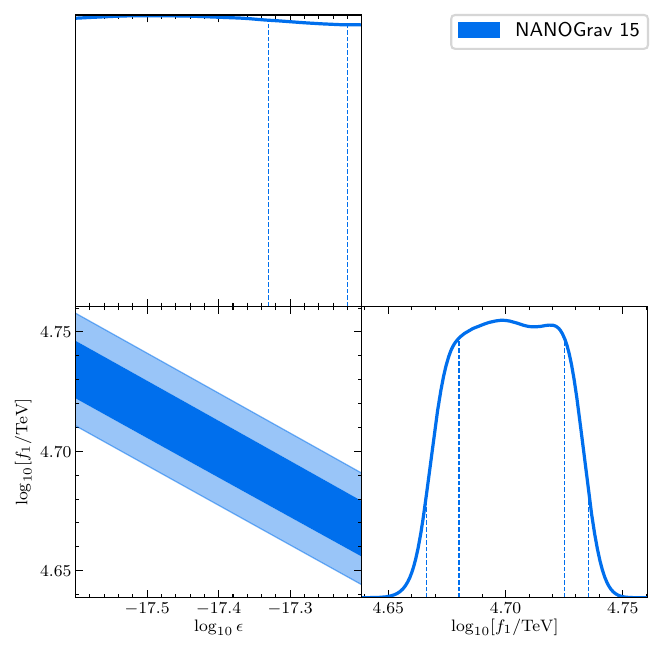}
    \caption{Constraints from NG15~\cite{NANOGrav:2023gor} on the parameters of GWs generated by $A_1$-wall annihilation in the minimal clockwork axion model. The best-fit values are $f_{1}=10^{4.7}$~TeV and $\epsilon=10^{-17.4}$.}
    \label{A_fig_02}
\end{figure}

In this appendix, we consider the opposite case: one can first analyze the GW signals generated by the annihilation of the \(A_1\)-wall via instanton effects, fitting these signals to the NG15 data, before subsequently examining the impact arising from the \(A_2\)-wall.
Similar to the analysis in Section~\ref{Section3}, as shown in Fig.~\ref{A_fig_01}, the yellow and green regions represent the $1 \sigma$ and $2 \sigma$ bounds obtained from fitting the NG15 data, respectively.
Figure~\ref{A_fig_02} illustrates the fitting results for the model parameters in the minimal clockwork model, with the best-fit values being $f_{1}=10^{4.7}$~TeV and $\epsilon=10^{-17.4}$.

\begin{figure}[htbp]
    \centering
    \includegraphics[width=0.95\linewidth]{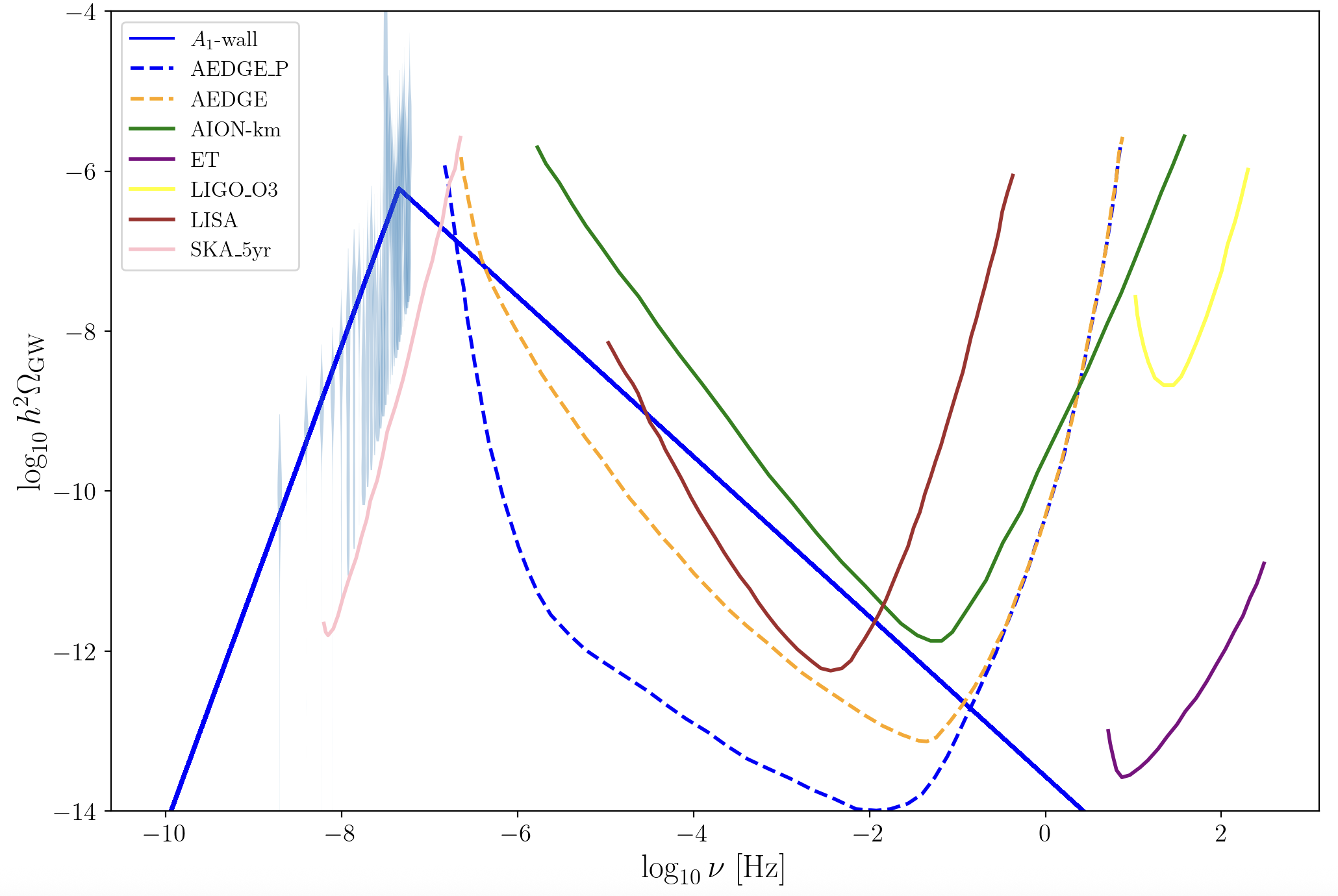}
    \caption{Spectrum of the GW signal (the blue solid line) induced by the annihilation of $A_1$-wall in the minimal clockwork axion model, where we take $f_{1}=10^{4.7}$~TeV, $\epsilon=10^{-17.4}$. Also shown are the sensitivity curves of SKA 5yr~\cite{Janssen:2014dka}, AION-km~\cite{Badurina:2019hst}, AEDGE~\cite{AEDGE:2019nxb}, AEDGE+, LISA~\cite{Caprini:2019egz}, LIGO O3~\cite{Shoemaker:2019bqt}, and Einstein telescope (ET)~\cite{Maggiore:2019uih}.  }
    \label{A_fig_03}
\end{figure}

\subsection{Consequences of $A_{2}$-wall}
Assuming in this case that the $A_2$-wall also annihilates via the instanton effect, applying Eqs.~(\ref{Tann_01}) and (\ref{Tdom}) yields $T_{\text{ann}} \simeq 1124.66~\mathrm{GeV}$ and $T_{\text{dom}} \simeq 0.031~\mathrm{MeV}$. 
However, at such a high annihilation temperature, the instanton effect is negligible, contradicting the assumption. Thus, the $A_2$-wall requires an alternative annihilation mechanism as well.
Considering a bias term from Eq.~(\ref{Gen_Tann}) with $V_{\text {bias }}^{1 / 4}=1200$~GeV, we obtain \(T_{\text{ann}} \simeq 576.5 \, \mathrm{GeV}\), \(\nu_{\text{peak}} \simeq 9.41 \times 10^{-5} \, \mathrm{Hz}\), and \(\Omega_{\text{GW}}^{\text{peak}} \simeq 3.07 \times 10^{-36}\). Unfortunately, this GW signal is exceedingly weak and remains undetectable by the currently foreseeable experimental facilities.
This very weak signal originates from the scaling relation $h^2\Omega_{\text{GW}}^{\text{peak}} \propto \epsilon \cdot f^6$, combined with the fact that $\epsilon$ is fixed at an extremely small value by the $A_1$-wall dynamics ($\epsilon = 10^{-17.4}$) and that $f_2 \ll f_1$.
As shown in Fig.\ref{A_fig_03}, the model predicts in this case only one detectable gravitational wave signal, which is associated with the \(A_1\)-wall and exhibits a peak frequency around \(10^{-7.5}\)~Hz.

\end{document}